\documentclass[sigconf,authorversion]{acmart}

\usepackage{multirow}
\usepackage{enumitem}

\usepackage{subcaption} 

\usepackage{graphicx}

\acmYear{2020}\copyrightyear{2020}
\setcopyright{acmcopyright}
\acmConference[WiSec '20]{13th ACM Conference on Security and Privacy in Wireless and Mobile Networks}{July 8--10, 2020}{Linz (Virtual Event), Austria}
\acmBooktitle{13th ACM Conference on Security and Privacy in Wireless and Mobile Networks (WiSec '20), July 8--10, 2020, Linz (Virtual Event), Austria}
\acmPrice{15.00}
\acmDOI{10.1145/3395351.3399421}
\acmISBN{978-1-4503-8006-5/20/07}

\begin{document}

\title{Peek-a-Boo: I see your smart home activities, even encrypted!}

\author{Abbas Acar$^1$, Hossein Fereidooni$^2$, Tigist Abera$^2$, Amit Kumar Sikder$^1$, Markus Miettinen$^2$, Hidayet Aksu$^1$, Mauro Conti$^3$, Ahmad-Reza Sadeghi$^2$, Selcuk Uluagac$^1$}

\affiliation{%
 \institution{$^1$Florida International University -  \{aacar001,asikd003,haksu,suluagac\}@fiu.edu}
}

\affiliation{%
 \institution{$^2$TU Darmstadt -  \{hossein.fereidooni,tigist.abera,markus.miettinen,ahmad.sadeghi\}@fiu.edu}
}

\affiliation{%
 \institution{$^3$University of Padua - conti@math.unipd.it}
}

\renewcommand{\shortauthors}{Acar et al.}

\begin{abstract}

A myriad of IoT devices such as bulbs, switches, speakers in a smart home environment 
allow users to easily control the physical world around them and facilitate their living styles through the sensors already embedded in these  devices. 
%
%
Sensor data contains a lot of sensitive information about the user and devices.
However, an attacker inside or near a smart home environment can potentially exploit the innate wireless medium used by these devices to exfiltrate sensitive information from the encrypted payload (i.e., sensor data) about the users and their activities, 
invading 
user privacy. 
With this in mind, in this work, we introduce a novel \textit{multi-stage privacy attack} 
against user privacy in a smart environment. 
It is realized utilizing 
state-of-the-art machine-learning approaches for detecting and identifying the types of IoT devices, their states, and ongoing user activities in a cascading style by only passively sniffing the 
network traffic from smart home devices and sensors.   
The attack effectively works on both encrypted and unencrypted communications.
We evaluate the efficiency of the attack with real measurements from an extensive set of popular off-the-shelf smart home IoT devices 
utilizing a set of diverse network protocols like WiFi, ZigBee, and BLE.
Our results show that  
an adversary passively sniffing the 
traffic can achieve very high accuracy (above 90\%) in identifying the state and actions of targeted smart home devices and their users. 
To protect against this
privacy leakage, we also propose a 
countermeasure 
based on generating spoofed 
traffic to hide the device states 
and demonstrate that it 
provides better protection than existing solutions. 

\end{abstract}



\maketitle

\section{Introduction}

Previously, the Internet was mainly used for accessing and displaying content of web pages (i.e., web browsing). However, with the emergence of IoT devices in smart homes, users have now the ability to control their home's electronic systems (e.g., smart bulbs, smart locks, sensors) using appropriate smartphone apps and also from remote locations~\cite{sikder2018survey, sikder2018iot}.  
To realize smart home automation, the devices are mostly equipped with embedded sensors. These sensors collect data from the environment and help users to control them. Moreover, smart home devices are also continuously communicating with associated back-end system servers or 
other devices (e.g., smart hubs) to transmit the sensor data in a real-time  manner. 
On the other hand, as IoT devices usually are single-purpose devices, the capabilities of 
individual smart home devices are relatively limited, comprising only a few states or actions. 
For example, a motion sensor allows a user to detect any movement in a physical space,  but the sensor has only two states: motion and no-motion. If an attacker can reveal the current state of the sensor, the attacker will also reveal the presence of the user at home.


Given that the communications among the server, sensors, smart-hub, and the smart home devices are usually encrypted using standard protocols like WPA2, in the case of WiFi, the contents of the exchanged messages or commands are hidden. 
However, the encryption only hides the payload, related meta-data (e.g., packet lengths, traffic rate) of the network traffic still leaks some information about the messages  exchanged~\cite{sun2002statistical,velan2015survey,stober2013you,li2016demographics,7265055,berkay2018SaintTaintAnalysisUsenixSecurity}. 

Identification of the encrypted 
traffic is a well-studied problem. However, applying traditional identification methods such as statistical techniques~\cite{sun2002statistical} 
in the domain of smart home is not straightforward due to challenges arising from the inherent properties of IoT devices. First, unlike targets using a widely-deployed protocol to perform a well-known specific activity like web browsing, in the smart home context, the targeted device population is much more heterogeneous and uses
various 
network protocols such as WiFi, ZigBee, BLE, etc. for supporting an even wider variety of device-type-specific, potentially proprietary application protocols. 
This naturally extends the potential attack surface, but also makes it even harder to devise \emph{generic} attacks or countermeasures.


Some earlier works~\cite{srinivasan2008protecting,apthorpe2017spying,apthorpe2017smart} have shown that it is relatively easy to make some simple inferences such as device type inference~\cite{miettinen2017iot}, identifying the user occupancy via detecting the mode transition between the device activities~\cite{copos2016anybody}, or simple device mode inference~\cite{apthorpe2017spying}. However, combining such partial information from different smart home devices to get a more meaningful picture about a user's actions or his/her activity profile is challenging. This is because a successful attacker must aggregate information about actions over a longer period of time from a multitude of smart home devices, which is only feasible if activity detection and identification can be \emph{automated} to a large degree to keep the required effort manageable.

In this paper, we demonstrate how machine learning methods based on traffic profiling of smart home IoT device communications can be used by an adversary to automatically identify actions and activities of the IoT devices and its users in a victim's smart home with very high accuracy, even if only encrypted data are available. Indeed, device types, daily mundane activities of the users (e.g., left home, walking from kitchen to bedroom), or states of the devices (e.g., door locked, unlocked) can all be easily identified even if the traffic is encrypted, thus posing a 
threat to user privacy.
We refer to this novel attack to user privacy as \textit{multi-stage privacy attack}, which is achieved in a cascading style by only observing passively the wireless traffic from smart home devices. In this, a passive attacker can easily realize the multi-stage privacy attack to extract meaningful data from any smart environment equipped with smart devices including personal homes, residences, hotel rooms, offices of corporations or government agencies.  
Here, unlike earlier approaches, the presented attack is device-type and protocol-agnostic, making it easily applicable to a wide variety of different IoT device types without the need for tedious harvesting of device-type or protocol-specific knowledge about specifications for supporting 
the activity identification task. 

We evaluate the effectiveness of the novel multi-stage privacy attack 
with 22 different off-the-shelf IoT devices utilizing the most popular wireless protocols for IoT. 
Our experimental results show that an attacker can achieve very high accuracy (above 90 \%) in identification of the types, actions, states, activities of the devices and sensors. 
Moreover, to 
counter 
the identified privacy threats posed by the multi-stage privacy attack, we also propose a new effective countermeasure solution based on generating spoofed traffic to hide the real states of targeted IoT devices and thereby the real activities of the users. Our solution does not require modifications in targeted IoT devices and is, therefore, easier to deploy than previously proposed solutions for IoT devices, for which it is very difficult to implement client-based countermeasures 
due to the vast heterogeneity of smart devices and limited resources available on 
the IoT devices. Also, even if the user is not at home,  a fake traffic-based solution for the user's presence will mask the user's absence, further improving privacy.  


\vspace{3pt}
\noindent\textbf{Contributions :} The contributions of this work are as follows:
\begin{itemize} [leftmargin=*,topsep=0pt] 
\item We propose a \emph{novel multi-stage privacy attack} on smart home users and devices which can leak sensitive information including types of devices, states of the devices, sensor data, and on-going user activities. To the best of our knowledge, this is the first work showing all the stages of an attack that can reveal the user activities from the raw 
    traffic of smart home devices, even encrypted.
    
\item The attack includes several novel techniques, both theoretical and practical, for effectively reducing the effort needed for the user activity inference on the timing-based heterogeneous network traffic. First, we demonstrate how to convert the user activity inference on the timing-based heterogeneous network traffic into a cascaded Machine Learning (ML) problem. Then, we further extend the attack by modelling the user activities via the Hidden Markov Model (HMM).
\item We evaluate our proposed novel attack with a dataset of 22 popular commercial smart home devices. We show that an attacker can automatically detect and identify device actions with high accuracy ($> 90\%$), allowing an adversary to infer potentially sensitive information about the smart home users.
\item Finally, although the focus of this paper is on the novel attack, we also propose a new solution based on traffic spoofing to address this new privacy threat while demonstrating its efficacy over existing solutions (Sec.~\ref{sec:solution}). 
\end{itemize}


\noindent\textbf{Organization}: The rest of this paper is organized as follows: In Section~\ref{sec:adversary}, we present the adversary model considered in this paper. Then, a background about 
the communication features of the smart home devices are presented in Section~\ref{sec:devices}. 
Section~\ref{sec:multi-stage} details the stages of our main proposed multi-stage attacks, where the results are presented in every sub-section. In addition, we present a solution to mitigate this privacy leaks in Section~\ref{sec:solution} and we discuss some related issues in Section~\ref{sec:discussion}. Finally, the related work are given in Section \ref{sec:related} and the paper is concluded in Section \ref{sec:conclusion}.

\section{Adversary Model} \label{sec:adversary}



\begin{figure} [h]
    \centering
    \includegraphics[width=.45\textwidth,trim=0cm 0cm 0cm 0cm]{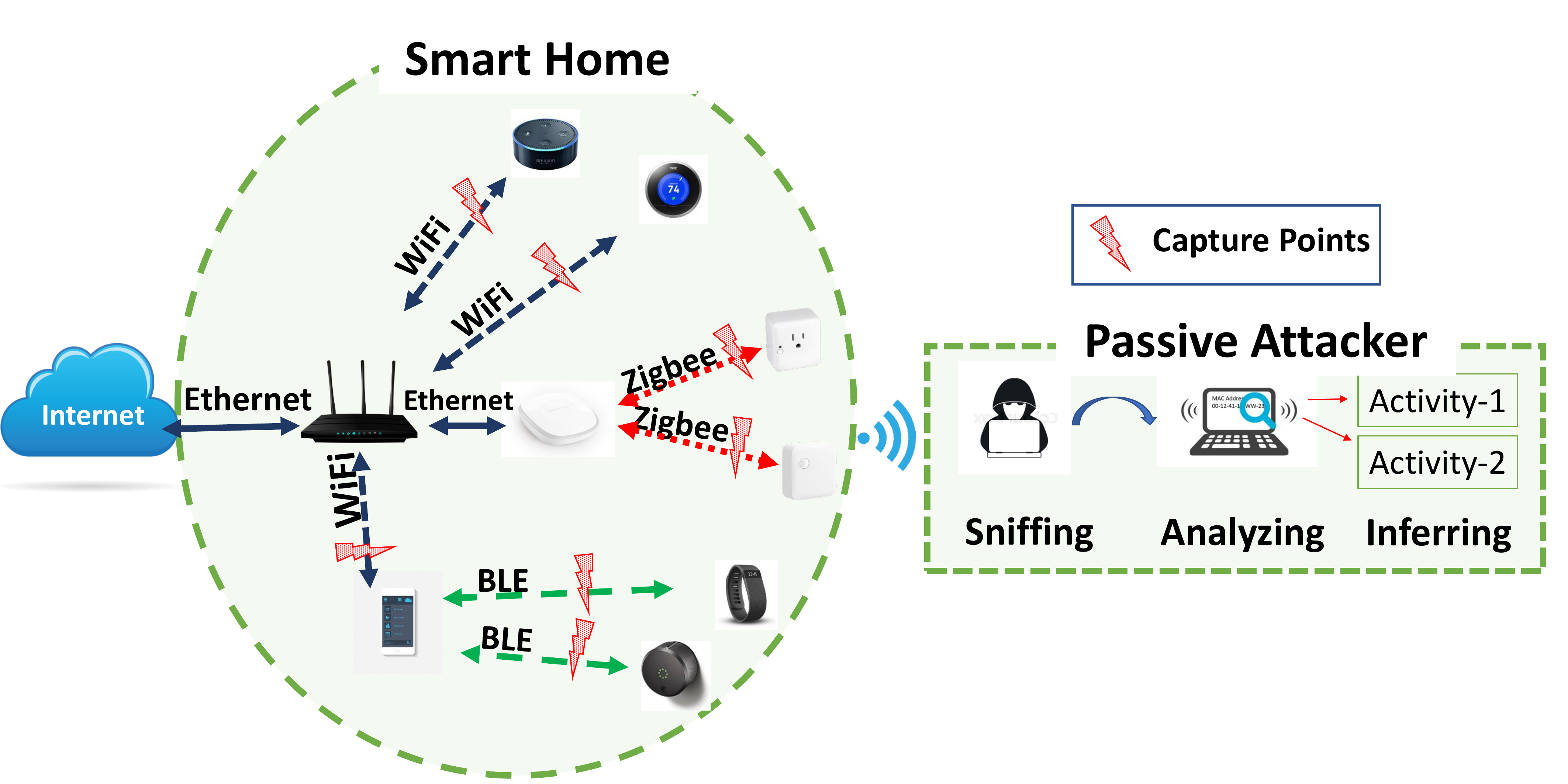}
     \caption{Local adversary model considered in this paper.} 
    \label{fig:adversary}
\end{figure}

One of the unique challenges in the domain of IoT, and particularly smart home, is that the attack surface is naturally extended and comprises 
a diverse set of devices and sensors deployed at the user's home.
Figure~\ref{fig:adversary} shows 
some of the 
data capturing points that an attacker can take advantage of when inferring 
user activities. In this work, we consider a local 
adversary located 
physically
within the wireless range of the targeted user's smart home devices similar to~\cite{fawaz2016protecting,formby2016s,han2018you}. For this, the attacker can install the sniffers only once and even manage them remotely. Or, it could compromise a device inside the smart home, remotely, and turn it into a sniffer. In this way, the attacker may never need to be present. In all these cases, the adversary can eavesdrop on various wireless IoT 
network communications transmitted by the user's smart home devices. 
For example, as presented in Figure~\ref{fig:adversary}, the attacker can sniff all the network traffic transmitted over WiFi, BLE, and ZigBee protocols. 
The attacker only needs to passively sniff the network traffic and does not need to interrupt.  
Therefore, the attacker may stay active long  enough 
without being detected by the victim.
An alternative adversary would be an adversary who can launch the attack remotely, i.e., intercepting the network traffic over the Internet such as a malicious ISP. We further discuss the advantages and limitations of such an adversary in Section~\ref{sec:discussion}.

\vspace{3pt}
\noindent \textbf{Assumptions.} We further make the following assumptions:
\begin{itemize} [leftmargin=*,topsep=0pt]
    \item 
    The attacker has access to the same kind of smart home devices and sensors as the targeted user, s/he can analyze the devices 
    by collecting 
    the 
    traffic of these devices, and use the collected data to train its algorithms. 
    \item 
    The attacker has access to protocol headers data on all layers that are not protected by encryption. The attacker does not need to know the specifications of analyzed protocols, instead it only needs to know how to run the already publicly available scripts, which does not require an extensive knowledge about the specifications of the protocol itself. Moreover, it can also use Layer 2
    information like MAC addresses, or BLE advertisement packets, to automatically identify additional information, the brand of individual devices, thereby reducing 
    the search space of devices to guess the set of smart home  devices that the targeted user is using. Moreover, it is also worth noting that the attacker does not need exactly same devices to train its algorithm, but it needs exact brand and device type to get the results presented in this paper as we use the $<brand,device-type>$ pair to uniquely identify devices.
\end{itemize}

\section{Smart Home Devices} \label{sec:devices}


In this section, 
we present the background information of the communication protocols used by the smart devices.

\begin{table}
    \centering
    \vspace{-5pt}
    \caption{The communication protocols and capabilities of the smart home devices used. 
          }
          
          \includegraphics[width=.48\textwidth,trim= 1.5cm 13cm 7.25cm 2cm]{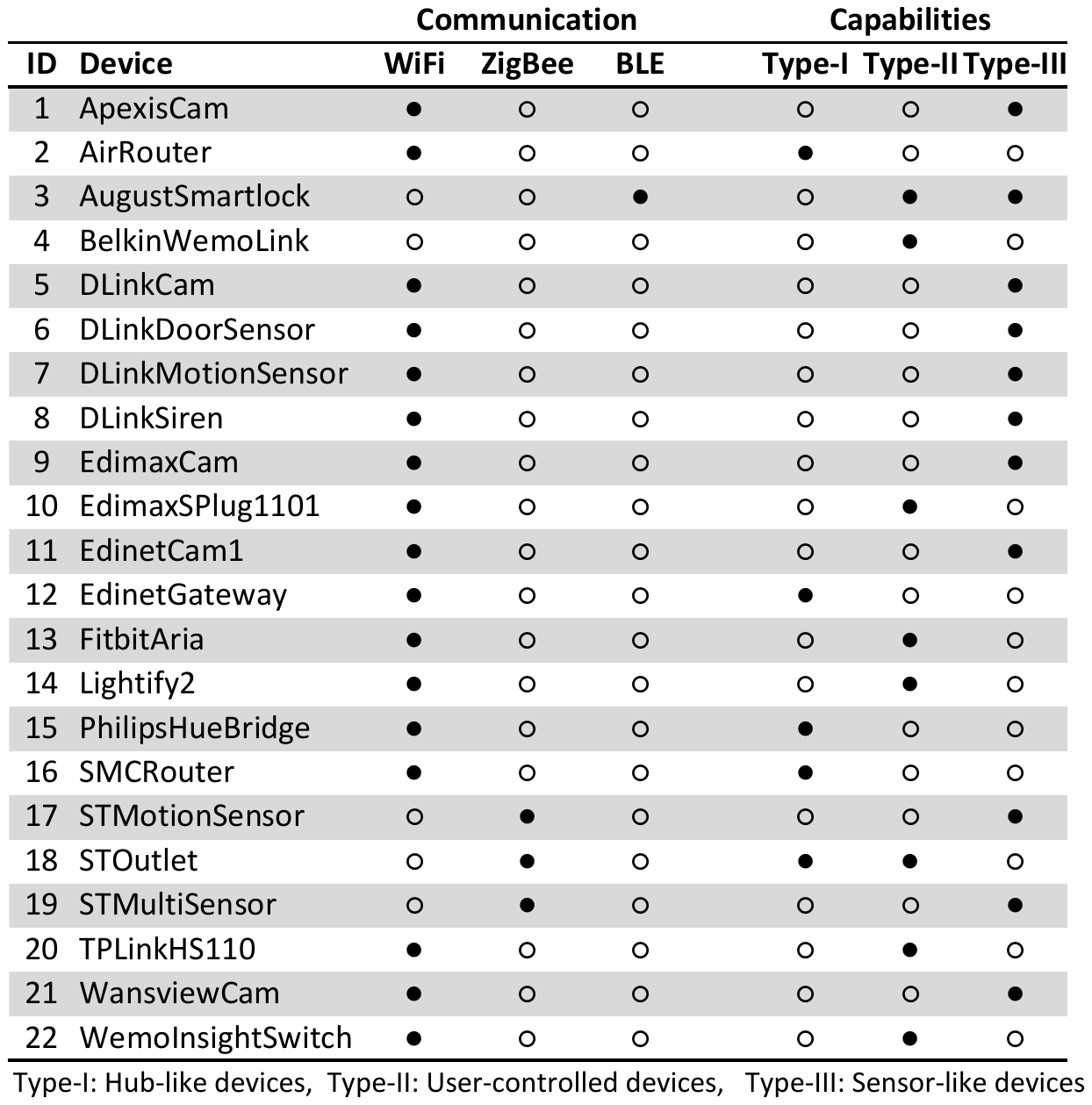}
          
\label{table:devices}
\vspace{-15pt}
\end{table}

\subsection{Communication Features}
Both the smart home vendors and users mostly prefer wireless communication over wired communication as it is more convenient. 
However, compared to wired communication, the wireless network traffic from smart home devices
is open to the eavesdropping attacks. 

In this work, we target three wireless protocols: WiFi, ZigBee, and Bluetooth Low Energy (BLE). Table~\ref{table:devices} shows  all the devices studied in this paper Table~\ref{table:devices} with their wireless communication protocol and device capabilities. 
Among these protocols, WiFi is used in the wired or plugged-in devices, while other protocols, ZigBee and BLE, are implemented for short range communication tasks of battery-powered devices as they consume less power than WiFi. 
\subsubsection{WiFi-enabled devices} 
WiFi-enabled devices are connected to the Internet either through a Hub-like device or directly connected to an access point. In both cases, the adversary can track and capture the traffic through a specific device via MAC address. Even though MAC addresses may help the attacker to narrow down the device type, it can not precisely decide the device type from MAC address. It may want to use IP addresses of servers. However, the adversary can only see the traffic that is encrypted by both the network protocols (SSL/TLS) and WiFi encryption (WPA). Therefore, it cannot see the IP or transport layer headers encrypted by the WPA protocol. This prevents the attacker from using header-based features for the device identification. However, the traffic rates of the devices still cannot be hidden from the attacker. 
\subsubsection{ZigBee-enabled devices} 
ZigBee devices have two addresses: MAC address and Network Address (NwkAddr). The MAC address is exactly the same as the MAC used in WiFi-enabled devices, which is unique for every device in the world and never changes. On the other hand, NwkAddr is created and assigned when the device joins a network and changes when it leaves and re-joins another network. It is similar to IP, however, it is not encrypted and source and destination NwkAddr of the packets can be seen by the attacker. In addition, the network coordinator (i.e., hub) has the $0x0000$ address and each network has a unique identifier, called the Personal Area Network Identifier (PAN ID). This information may additionally help the attacker.

\subsubsection{BLE-enabled devices}
In a BLE network, a device can be either a master or a slave. A slave can connect to only one master node while a master can connect to multiple slave nodes. In all the smart home devices that we used, while the smartphone acts as a master, targeted smart device acted as a slave. Before establishing the connection, a slave device broadcasts advertising packets (ADV\_IND) randomly on channel 37, 38, and 39. Once a connection starts, they agree on a channel map, where they follow in the rest of the communication. If an attacker wants to follow the BLE traffic through a smart device, it needs to capture the first packet so that it can learn the channel mapping. Once the attacker captures the access address, it can follow the rest of the communication. 


\section{Multi-stage Privacy Attack} \label{sec:multi-stage}
\begin{figure*}
    \centering
    \includegraphics[width=.80\textwidth,trim=0cm 1cm 0cm 1cm]{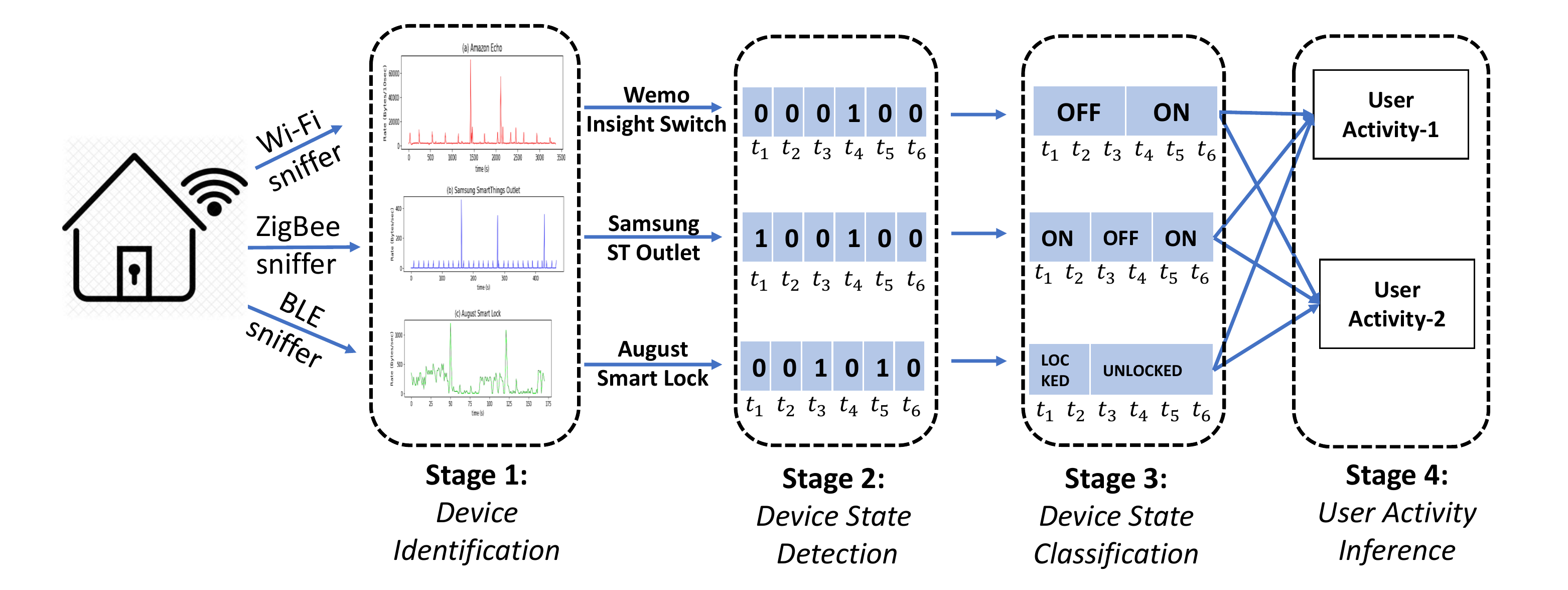} 
     \caption{Overview of our multi-stage privacy attack.
     } 
    \vspace{-10pt}
    \label{fig:multi-stage}
\end{figure*}
As shown in Figure~\ref{fig:multi-stage}, our novel multi-stage privacy attack consists of four stages connected in a cascaded manner. While the goal of the attack is to infer user activities at the final stage, every stage also leaks partial information about devices and their actions and can be independently used by the attacker for various purposes. 
Before going into further details about our attack, in Appendix~\ref{sec:case_studies}, we show the feasibility and possibility of privacy leaks from encrypted network traffic of smart home devices. Particularly, we show that an attacker who can sniff the network traffic of the devices can easily infer some simple information without using any advanced techniques.  We consider one device for each protocol: 
Wemo Insight Switch (WiFi), 
Samsung ST Outlet (ZigBee), and August Smart Lock (BLE). We analyze the raw network traffic of each device and see if it is really possible to extract information from the network traffic, specifically from data rate. In the next sections, we describe the details of individual stages and evaluate the efficiency of our multi-stage privacy attack on network traffic data collected from 22 different off-the-shelf IoT devices used in smart homes.

\subsection{Dataset and Evaluation Metrics}
In order to evaluate the attacks in the stages above, we collected the network data from 22 different smart home devices. Data collection was performed in two stages: In the first stage, controlled experiments were performed in which 
detailed instructions were followed to initiate specific actions on the tested device. These instructions were compiled based on the on-line or hardcopy manual of each tested device (specs and data sheets). The controlled experiments were performed in order to ensure that all relevant actions for each device were represented in the usage dataset sufficiently many times. Each experiment was therefore repeated $n=20$ times for each device. In addition to the controlled experiments, also uncontrolled testing was performed in order to capture background traffic of relevant devices. In this set-up, several devices were configured to be used simultaneously and device actions were 
occasionally triggered 
during a test period of ca. 1-2 hours.

The duration and the total size of the captures and the number of the packets are given in Table~\ref{table:dataset}. 
The devices used include 
a representative cross-section of IoT device types,  typically available in the European and North American markets during the study.
The devices were also selected based on the market share of different device categories. The most popular device categories are smart security systems such as smart cameras and smart locks (22.2\%), lighting (3.03\%), outlets and switches (1\%), gateways including hubs and routers (24.5\%), and smart speakers (22.39\%)~\cite{survey}. In addition to these categories, we also included several smart sensors as these devices hold significant smart home market share (approximately 23.9\%)~\cite{sensor}. We installed all the devices in a laboratory network and emulated user inputs triggering device state changes. We captured all the network traffic from a device and performed the analysis offline. 


For evaluating the efficiency of our attacks, we  use different metrics. 
First, we use accuracy, which is the ratio of correctly inferred observations to total observations. In some cases, as in real deployments, the collected network data may have imbalanced data, where the duration of the active state is much less than the inactive one. In those cases, we use additional metrics such as Precision, Recall, F1 score, and Support. In the cases that the dataset includes a lot more label  0 (no activity) rows than label 1 (activity) rows, we observed that F1 score is a better performance measurement than accuracy although accuracy is a more intuitive performance measurement, in general. The detailed calculation of these evaluation metrics is given in Appendix~\ref{sec:metrics}.


\begin{table}

\caption{Characteristics of network traces used in experiments.
}
\vspace{-5pt}
\label{table:dataset}
\centering
\resizebox{.85\columnwidth}{!}{%
\begin{tabular}{@{}|l|c|c|c|@{}}
\hline

 \textbf{Device} &  \textbf{Period (mins)} &  \textbf{Size (MB)}&  \textbf{Packets} \\
\hline
ApexisCam           &   133   & 80    & 152220  \\ \hline
AirRouter               &   85    & 49    & 115192   \\ \hline
AugustSmartLock             &    25.8   &   0.66  &  8129  \\ \hline
BelkinWemoLink         &   71    & 0.66  & 2039    \\ \hline
DLinkCam            &   225   & 1.15  & 5389    \\ \hline
DLinkDoorSensor       &   74    & 0.48  & 3519    \\ \hline
DLinkMotionSensor     &   74    & 0.47  & 2849    \\ \hline
DLinkSiren             &   71    & 0.41  & 3073    \\ \hline
EdimaxCam           &   225   & 0.27  & 1798     \\ \hline
EdimaxSPlug1101        &   74    & 0.5   & 2823    \\ \hline
EdinetCam1& 117 &0.3& 2779 \\\hline
EdinetGateway& 225 &0.34& 3240  \\\hline
FitbitAria             &   213   & 0.043 & 257     \\ \hline
Lightify2               &   74    & 0.25  & 1022     \\ \hline
PhilipsHueBridge      &   53    & 0.8   & 2680    \\ \hline
SMCRouter              &   124   & 47    & 150768  \\ \hline 
STOutlet     &       6    & 0.04      &     1061      \\ \hline
STMotionSensor     & 11       & 0.05            & 1291       \\ \hline
STMultiPurpose    & 12            & 0.22            & 5255      \\ \hline
TPLinkHS110    &      71  &      0.14       &  473     \\ \hline
WansviewCam         &   193   & 11    & 73759   \\ \hline
WemoInsightSwitch&  117& 0.8&1675 \\  \hline
\end{tabular}
}
\end{table}

\subsection{Calculating Features from Network traffic}

In this sub-section, we explain how we use the traffic flow for the classification task. Particularly, we 
take advantage of the fact that while the 
encryption layer in the protocol protects the payload of a packet, it fails to hide other information revealed by network traffic patterns, for instance, sequence of packet lengths (SPL) and direction (incoming/outgoing). We consider each network traffic flow as a time ordered sequence of packets exchanged between two peers during a session. Before processing the network traffic for classification, 
we converted packet in traffic flow into a Sequence of Packet Lengths and Times (SPLT) as in following format: 
\begin{equation}
pkt= [timestamp,\ direction, \ packet \ length], 
\end{equation}

\noindent where the direction is 1(0) if it is an incoming (outgoing) packet. This transformation is done for each packet in the captured trace, where each result is written to a new row. In the end, we obtained a matrix with three columns. Then, in the feature extraction of each attack, we calculated the features from this matrix.


\subsection{Stage-1: Device Identification \label{subsec:type}}


Several different identification approaches for IoT devices have been
proposed in literature. 
Numerous works have shown 
that IoT devices can be identified 
with high accuracy for both WiFi-enabled ~\cite{miettinen2017iot,dalai2017wdtf,bezawada2018iotsense,meidan2017detection,nguyen2018d} and BLE-enabled~\cite{das2016uncovering} devices. 
Therefore, in this section (e.g., Stage-1), we implemented already existing device identification algorithm for ZigBee-enabled smart home devices using our features to see whether we can identify the ZigBee-enabled smart home devices from their network traffic.



In our dataset, each device can be uniquely identified by the $<brand, device-type>$ pair. We did not consider the different models of devices as different devices. On the other hand, a hub in ZigBee always uses the network address $0x0000$, so it can be easily recognized by the attacker. Therefore, we did not include the hub in the identification of ZigBee devices. After collecting ZigBee network traffic, the second step involves extracting the features to identify the devices. In this step, the features we used include \textit{mean packet length}, \textit{mean inter-arrival time}, and \textit{standard deviation in packet lengths}. We split each individual network traffic trace of a device into equal time intervals (e.g., 5 sec, 10 sec). Then, we calculated these features for each interval. 

For the classification, we used the kNN classification algorithm. 
The classifier could correctly identify devices with an overall accuracy of 
93\% 
for ZigBee devices. This shows 
that as for WiFi and BLE, also devices using ZigBee can be identified with high accuracy.

\subsection{Stage-2: Device State Detection \label{subsec:detection}}
When an interaction between the device and the user occurs, a significant amount of data is transmitted, which leads to a significant increase in the traffic rate. After this data exchange, the data transmission drops to the minimum until a new interaction starts. When there is no activity, only the minimum amount of continuation packets like heartbeat messages are sent to minimize the device's power and bandwidth consumption. We also observed that almost the same amount of data transfer occurs for the same activities. All this information allows us to detect transitions between the activities or states of the device. For further validation, we do the following experiments.

\begin{table}
\centering
\caption{Evaluation results of device activity detection stage.
}
\resizebox{.9\columnwidth}{!}{%
\begin{tabular}{|l|c|c|c|c|}
\hline
\multirow{2}{*}{\textbf{Device}}& \multicolumn{2}{c|}{ \textbf{Random Forest}} &\multicolumn{2}{c|}{\textbf{kNN}}\\ \cline{2-5}
 &F1 Score& Accuracy& F1 Score&  Accuracy\\
\hline
ApexisCam & 93 & 97 & 94 & 98\\ \hline
AirRouter & 98 &97&98&97 \\ \hline
AugustSmartLock  & 100 & 100 & 100   & 100  \\ \hline
BelkinWemoLink & 80 & 79 & 85 & 83\\ \hline
DLinkCam& 85 &80&85&80 \\ \hline
DLinkDoorSensor &94 & 98 & 92 &97\\\hline
DlinkMotionSensor & 74 & 96 & 69 &95 \\ \hline
DlinkSiren &89 & 99 &91 &99 \\ \hline
EdimaxCam& 84 &82&82&81 \\\hline
EdimaxSPlug1101& 91 &97&92&97 \\\hline
EdinetCam1& 76 &96&76&96 \\\hline
EdinetGateway& 80 &99&66&99 \\\hline
FitbitAria  & 100 & 100&100&100 \\ \hline
Lightify2& 86 &99&81&98 \\\hline
PhilipsHueBridge& 74 &98&76&98 \\\hline
SMCRouter& 94 &91&100&100 \\\hline
STOutlet      & 83 &     99& 92 & 99    \\ \hline
STMotionSensor  &91 & 97  & 92   &97  \\ \hline
STMultiSensor  &86 & 99 & 92   & 99    \\ \hline
TPLinkPlug1101& 98 &99&92&99 \\\hline
WansviewCam & 91 & 87 & 91 & 86 \\ \hline
WemoInsightSwitch& 86 &98&88&98 \\\hline

\textbf{Avg}& \textbf{88} & \textbf{99} &\textbf{91} & \textbf{95}  \\   \hline     
\end{tabular}
}
\label{tbl:perf}
\vspace{-.5cm}
\end{table}

\subsubsection{Feature Extraction}
Our goal is to transform a sequence of packets into a supervised learning dataset. To achieve this, we divided the sequence of packets into windows of size $W$. For a given time interval length $W$, we extracted a feature vector comprised of three variables: \textit{mean packet length}, \textit{mean inter-arrival time} and \textit{median absolute deviation of packet size}. Based on timestamped labels telling whether an activity was ongoing or not, we labeled the given vector with 1 for an ongoing activity or 0 for no activity.
We found that the window size has significant influence on the performance of our model. The window size for the best performance depends on adjusting the size according to the duration of the activity. In general, selecting a smaller window size improves the performance until some level, but any further reduction results in decline of the performance. From our observation, better performance was observed when the window size is about a quarter of the duration of an activity.
\subsubsection{Results}
After obtaining feature vectors with labels from the sequence of packets, any supervised learning algorithm can be applied on the dataset. We have evaluated two supervised learning algorithms, namely Random Forest classifier (RF) and k-Nearest Neighbors classifier (kNN). As shown in Table~\ref{tbl:perf} both RF and kNN have similar performance with RF averaging 88\% and kNN with 91\% average of correctly detecting activities.  F1 Score of each device in Table~\ref{tbl:perf} differs slightly. DlinkMotionSensor has the worst F1 score 74\% using RF and 69\% using kNN and the best F1 score is 100\% for the Aria Fitbit and AugustSmartLock.

\subsection{Stage-3: Device State Classification \label{subsec:classification}}
In the device state classification experiments, the attacker's goal is to decide the state of the device (e.g., deciding if it is ON or OFF).
When looking at the device's exchanged network packets, unlike previous steps, this is more difficult to determine.
However, each state has a unique pattern which helps us to differentiate them from each other. In order to see if it is possible to differentiate the states, we did the following experiments:  

\subsubsection{Feature extraction} To conduct device state classification, informative and distinctive features must be extracted from time-series generated in the preprocessing steps. We used the \textit{tsfresh}~\cite{tsfresh} tool that automatically calculates a large number of time series characteristics and features and then constructed our feature vector.  Examples of the features extracted from time-series are as follows:  Absolute Energy of time-series, Length of time-series, Mean and median of time-series, Skewness of time-series, Entropy of time-series, Standard deviation of time-series, Variance of time-series, Continuous wavelet transform coefficients, Fast Fourier Transform Coefficients, Coefficients of polynomial fitted to time-series. 

\vspace{-0.1cm}
\subsubsection{Feature selection} The output of the feature extraction phase is a set of feature vectors including 795 binary features. A large number of features, some of which redundant or irrelevant might present several problems such as misleading the learning algorithm, and increasing model complexity. A feature selection technique was therefore used to mitigate these problems and also to reduce over-fitting, training time and improve accuracy. We used a technique leveraging ensembles of randomized decision trees (i.e., Extra Trees-Classifier) for determining the importance of individual features. We exploited Extra-Trees Classifier to compute the relative importance of each attribute to inform feature selection. The features considered unimportant were discarded. The feature selection phase effectively reduced the feature vector size from 795 to 197 binary features. 

\subsubsection{Results}
Our objective was to build a 
performant model to correctly classify IoT devices' states even if their traffic is encrypted. To this end, we employed several machine learning algorithms for the classification such as \textit{XGBoost}, \textit{Adaboost}, \textit{Random Forest}, \textit{SVM with RBF kernel}, \textit{kNN}, \textit{Logistic Regression}, \textit{Na\"ive Bayes}, and \textit{Decision Tree}. 
In order to ensure that our machine learning model got 
the most of the patterns from the training data correctly, and it was not picking up too much noise, we shuffled and split the data-points to conduct the following experiments: \textit{(i)} we performed 5-fold Cross Validation (CV) on a training set of 377 samples (75\% of data) for assessing the effectiveness of the machine learning model and \textit{(ii)} we carried out Hold-out Validation on 126 samples (25\% of data) to test the machine learning model performance against unseen data.


\begin{table}
\centering
\caption{Cross-validation and hold-out validation results for device state classification. }
\vspace{-5pt}
\label{my-label}
\resizebox{.85\columnwidth}{!}{%
\begin{tabular}{|l|c|c|c|c|}
\hline
\multirow{2}{*}{\textbf{Classifier}} & \textbf{5-fold CV} &   \multicolumn{3}{c}{\textbf{Held-out data} (25\% of data)} \\ \cline{3-5} 
 & (75\% of data)  & Precision   & Recall  & F1 Score   \\ \hline
SVC RBF Kernel              & 86                                        & 89              & 87            & 87            \\ \hline
Logistic Reg.               & 87                                      & 90               & 89            & 88            \\ \hline
\textbf{Random Forest}               &  \textbf{92}                                       & 96              & 94           & \textbf{94 }            \\ \hline
Naive Bayes                 & 87                                       & 92             & 87            & 88            \\ \hline
Decision Tree               & 66                                       & 62              & 63            & 61           \\ \hline
K-NN                        & 84                                        & 91              & 87            & 87             \\ \hline
Adaboost                    & 86                                        & 89              & 87            & 87             \\ \hline
XGBoost                     & 85                                        & 91              & 87           & 87            \\ \hline
\end{tabular}
}

\label{cv}
\end{table}


\textbf{5-fold Cross Validation}:
To avoid the risk of missing important patterns or trends in the dataset, we applied
cross validation, as it provides ample data for training the model and also leaves ample data for validation. Thus, we conducted a 5-fold cross validation experiment. In 5-fold CV, the data are randomly partitioned into 5 equal-sized sub-samples. Of the 5 sub-samples, a single sub-sample is retained as the validation data for testing the model, and the remaining 4 sub-samples are used as training data. The process is then repeated 5 times with each of the 5 sub-samples used exactly once as the validation data. The 5 results from the folds can then be averaged to produce a single estimation. We obtained 92\% accuracy in terms of F1 Score in the detection of devices' states using Random Forest classifier, as shown in Table ~\ref{cv}.

\textbf{Hold-out Validation}: To make sure that our classifier can generalize well and is not over-fitted, we tested the classifiers' performance in terms of Precision, Recall, and F1 Score against unseen data (the data was removed 
from the training set and is only used for this purpose). Table ~\ref{results} shows the detailed results obtained by Random Forest classification algorithm when conducting the device state classification over 126 unseen samples. As can be seen, the F1 Score of each device used in the experiment differs slightly. 
We obtained an average performance measurement of 0.94 (94\%) of correctly classifying activities. This shows that an attacker can easily differentiate the devices' states.

\begin{table}
\caption{Hold-out 
validation results of RF classifier for all IoT devices. }
\vspace{-5pt}
\label{results}
\resizebox{.9\columnwidth}{!}{%
\footnotesize
\begin{tabular}{|l|l|c|c|c|c|}
\hline
\textbf{Device name} & \textbf{Action} & \textbf{Pre.} & \textbf{Recall} &\textbf{F1}& \textbf{Supp.}\\\hline
ApexisCamera & live view & 100 &100  & 100&4 \\\hline
AirRouter   & surfing on amazon & 80 & 100 & 89& 4 \\\hline
AugustSmartLock& off & 100 & 67 &80 & 3 \\\hline
AugustSmartLock& on & 67 & 100 &80 & 2 \\\hline
BelkinWemoLink     & off & 80 &100  &89 & 8 \\\hline
BelkinWemoLink     & on & 100 &50 &67 & 4 \\\hline
DLinkCamera       &  live view & 100 & 100 &100 &3  \\\hline
DLinkDoorSensor    & open & 100 & 100 &100 &5  \\\hline
DLinkSensor        & motion detection  & 100 & 100 & 100& 6 \\\hline
DLinkSiren         & turn on & 100 & 100 &100 &1 \\\hline
EdimaxCam       & live view  & 100 & 100 & 100& 1 \\\hline
EdimaxSPlug1101         & on & 100 & 100 & 100&5  \\\hline
EdinetCam1       & live view  & 100 & 100 &100 &2  \\\hline
EdinetGateway       & on  & 100 &100  &100 &3  \\\hline
FitbitAria &  measure weight & 100 &100  & 100&4  \\\hline
Lightify2 & change light type & 100 & 100 &100 & 6 \\\hline
PhilipsHueBridge & turn scene off & 100 & 100 &100 &3  \\ \hline
PhilipsHueBridge & turn scene on & 100 & 100 &100 &5  \\\hline
SMCRouter   & surfing on amazon  & 100 & 80 &89 &5  \\\hline
STOutlet   & on & 100 & 89 &94 & 9 \\\hline
STMotion     & active &88  & 100 &93 & 7 \\\hline
STMotion     & inactive &100 & 71 &83 & 7\\\hline
STMultiSensor  & acceleration active &100  & 100 & 100& 8 \\\hline
STMultiSensor  & acceleration inactive &71  & 100 & 83& 5 \\\hline
TPLinkPlugHS110     &turn off  &100  & 100 &100 &5  \\\hline
WansviewCam     &reboot   & 100 & 100 & 100& 9 \\\hline
WemoInsightSwitch  & on & 100 & 100 &100 &  2\\\hline
 \textbf{Avg./Total}  & ----------- & \textbf{96} & \textbf{94} &\textbf{94} & \textbf{126}  \\             
\hline
\end{tabular}
}
\vspace{-15pt}
\end{table}

\subsection{Stage-4: User Activity Inference \label{subsec:activity} }
Modern smart home environments comprise several sensors and devices that are connected with each other and share information. These devices and sensors are configured as independent entities, but work co-dependently to provide an autonomous system. Any user activity in a smart home can be inferred by observing the states of the devices and sensors~\cite{sikder2019context, sikder20176thsense}.

\subsubsection{Modelling User Activities} 
\begin{figure}
  \centering
    \includegraphics[width=\columnwidth]{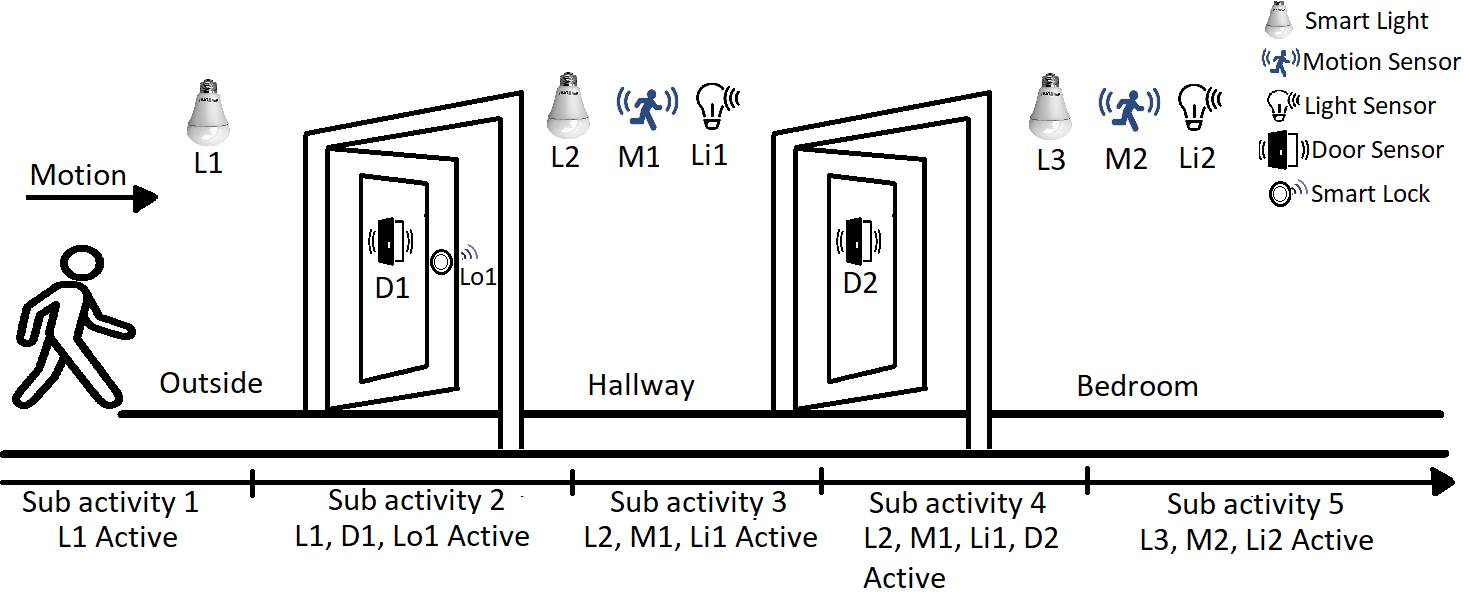}
     \caption{User walking scenario in a smart home.}
     \label{context}
\end{figure}

In Figure~\ref{context}, we demonstrate a simple walking scenario of a user. Here, a user is entering the smart home from outside to the bedroom through the hallway. The scenario consists of five different devices with lights both inside and outside the home controlled by the motion sensor (M) and light sensor (L). This simple activity can be illustrated as a sequential pattern: Sub-activity 1- moving towards the door from outside (L1 is active), Sub-activity 2- user opens the front door (L1, D1, Lo1 are active), Sub-activity 3- user enters the hallway (L2, M1, Li1 are active), Sub-activity 4- user enters the room (Li2, L2, M2, D1, Lo1 are active), Sub-activity 5- user inside the home (L2, M2, Li2 are active). To complete the activity, a user must follow the same sequence of sub-activities and complete each step. Here, the devices' states (active/inactive) for a specific time can be determined from the network traffic captured from the devices. These device states can be used to infer an on-going activity in a smart home setting. 

\subsubsection{Feature Extraction} To infer user activities, different device features must be extracted from network traffic data. Network traffic data contain several features including timing information, sensor information, device states, location, etc. Based on the data-type, the extracted features from the network traffic for user activity inference can be represented as follows:
\begin{equation}\label{eq1}
Data\ array, E_T = \{S, D, M, L\},
\end{equation}
where T is the timing features extracted from the network traffic, S is the set of sensors' features, D is the set of device features, M is the features extracted from the controlling device (smartphone/tablet), and L is the set of location features extracted from the network traffic. We describe the characteristics of these features below.
\begin{itemize}[wide=0pt]
    \item \textit{Timing features (T):} Smart home devices change their state according to user activities and commands. Some devices perform time-independent tasks (e.g., switching lights with motion), while some devices perform a task in a certain pattern with different user activities (e.g., walking from one point to another) based on smart home settings. We extract the time of an event from the network traffic captured from different devices to build the overall state of the smart home at the time of the user activity.
    \item \textit{Sensor State features (S):} Smart home environment consists of different sensors (e.g., motion sensor, light sensor, door sensor, etc.) which act as a bridge between devices and the peripheral. Sensors in a smart home can sense different environment parameters which can trigger different pre-defined tasks in multiple devices. Moreover, sensors can sense any change occurred because of a user interaction and forward this information as an input to the associated devices. These sensor data can be both logical (motion sensor) and numerical (temperature sensor) depending on the nature of the sensor. We observe the changes in both logical and numerical value of a sensor from the captured network traffic and use as a feature to infer user activities. We represent the changes in sensor data as binary output: 1 for active state and 0 for inactive state.
    \item \textit{Device State features (D):} In a smart home environment, multiple devices such as smart light, smart thermostat, etc. can be connected with each other and with a central hub to perform different tasks. These devices can be configured to change their states (active/inactive) to perform a pre-defined task or to perform a task based on user activities. We consider the state information of all the connected devices as features and extract this information from captured network traffic to infer the on-going user activity. The active and inactive states of the devices are illustrated as 1 and 0 respectively in the data array.
    \item \textit{Controller State features (M):} Smart home devices can be controlled in an autonomous way and also by using a controller device (smartphone/tablet). To understand the changes in states of the sensors and devices, one should consider the control commands generated by the controller devices. We consider the state of controller device as active (represented as 1 in data array) when a user interacts with smart home devices via controller device and inactive otherwise (represented as 0 in data array). This state information of the controller devices can be extracted from the captured network traffic to build the data array.
    \item \textit{Controller Location features (L):} The devices connected in a smart environment can be controlled from a different location and this location information can be collected from the captured network traffic. We consider the location of the controller device as a feature to understand any activities on smart home. We consider the home location of the controller device as 1 and the away location of the controller as 0 to represent the location feature as a binary number in the data array.
\end{itemize}

\vspace{-0.1cm}
For Stage 4, we captured the network traffic from a smart home environment and create the feature array explained in Equation~\ref{eq1}. We captured the network traffic for a specific time to correctly portray user activities from the network data. Each element of the data array represents the operating conditions of different smart devices, sensors, and controller devices. These data were then used to train a Hidden Markov Model (HMM) to detect user activities in a smart home environment. 


To train this HMM, we collected data from a smart home environment with real smart devices. We consider common smart home devices to build our training environment~\cite{devices}. Our test smart home environment included Samsung SmartThings hub, Samsung multipurpose sensor, Samsung motion sensor, Netgear Arlo security camera, Philips Hue smart light, Ecobee Smart Thermostat, and August Smart Lock. We collected network traffic data from 10 different users for different user activities. 

\subsubsection{Activity Types}

\begin{table}[t]
\centering
\caption{Typical user activities in a smart home.}
\vspace{-5pt}
\resizebox{.9\columnwidth}{!}{%
\begin{tabular}{|l|l|}
\hline
\textbf{Task Category}                            & \textbf{Task Name}   \\ \hline

\multirow{3}{*}{Time-independent}      & 1. Controlling device within smart home.  \\\cline{2-2}
                                         & 2. Controlling device from outside of the home.\\\cline{2-2}
                                         & 3. Presence in a specific point at home.            \\ \hline
\multirow{3}{*}{Time-dependent} & 4. Walking in the smart home.       \\ \cline{2-2} 
                                         & 5. Opening/ closing doors/windows.               \\ \cline{2-2}
                                         & 6. Entering/ exiting from smart home \\ \hline
\end{tabular}
}

\vspace{-1pt}
\label{tasklist}
\end{table}

User activities in a smart home environment can be instantaneous (e.g., switching on a device) or sequential over time (e.g., walking from one place to another). We categorized user activities in a smart home environment in two categories - time-independent and time-dependent user activities. 

\begin{itemize}[leftmargin=*]
    \item \textit{Time-independent Activities:} These user activities are instantaneous, non-sequential activities which do not depend on time. For example, a user can switch on/off a device in the smart home environment at a specific time instance. This activity will show changes in different features for only one time. 
    \item \textit{Time-dependent Activities:} These user activities are time-dependent, sequential activities. For example, a user can move from one point to another point. This activity will show changes in different features over time in a specific sequence. 
\end{itemize}
We tested our HMM model with data collected from six different user activities. We selected these activities as these are the common user activities listed in prior works~\cite{sikder2019aegis}. 
As we intend to make our activity detection generalized, we do not consider any rare events that has less possibility of happening in a real-life environment. 
Our user activity model is explained below.
\begin{itemize}[leftmargin=*,topsep=0pt,noitemsep]
    \item \textit{User Activity- 1.} A user is controlling a device from inside of the smart home environment.
    \item \textit{User Activity- 2.} A user is controlling a device from outside of the smart home environment.
    \item \textit{User Activity- 3.} A user is performing tasks from a specific point of a smart home environment.
    \item \textit{User Activity- 4.} A user is walking from one point to another inside the smart home environment.
    \item \textit{User Activity- 5.} A user is entering/ exiting from the smart home environment.
    \item \textit{User Activity- 6.} A user is opening/ closing a window/ door in smart home environment.
\end{itemize}



\subsubsection{Results}

To train our proposed HMM for user activity inference, we collected user activity data for a week from 15 different people (total 30 datasets) in an emulated smart home environment. We asked the users to perform their daily activities in a timely manner (from morning to night) and performed the same activities in defined sequences in a real-life smart home setting. We considered single authorized smart home user interacting with smart devices at a time for data collection. We trained our HMM model with these data. We also collected data for this activity model to test our proposed method. We collected two datasets for each activity (12 in total) to test the efficacy of the activity inference model.

\begin{table}
\caption{User activity inference from network traffic data in a smart home environment.}
\label{performance}
\centering
\resizebox{.85\columnwidth}{!}{%
\begin{tabular}{|c|c|c|c|c|c|c|}
\hline
 Smart Home & \multirow{2}{*}{\textbf{TPR}} & \multirow{2}{*}{\textbf{FNR}} & \multirow{2}{*}{\textbf{TNR}}& \multirow{2}{*}{\textbf{FPR}} & \multirow{2}{*}{\textbf{Accuracy}} & \multirow{2}{*}{\textbf{F-score}} \\ 
 User Activity & & & & & &  \\ \hline
Activity-1    & 1      & 0      & 1         & 0     & 1
& 1  \\ \hline
Activity-2     & 1      & 0      & 1         & 0     & 1   & 1  \\ \hline
Activity-3             & 1      & 0      & 1         & 0     & 1 & 1  \\ \hline
Activity-4 & 0.96      & 0.03      & 0.94         & 0.05    & 0.95 & 0.95  \\ \hline
Activity-5 & 0.95     & 0.04      & 0.87         & 0.12     & 0.93 & 0.91  \\ \hline
Activity-6 & 0.97     & 0.02      & 0.91         & 0.08     & 0.94 & 0.94  \\ \hline
\end{tabular}
}
\vspace{-10pt}
\end{table}


In Table~\ref{performance}, the evaluation results of our activity inference model are shown. For time-independent activities (Activity-1, Activity-2, and Activity-3), one can infer with 100\% accuracy and F-score from the captured network traffic data in a smart home environment. On the contrary, accuracy and F-score decreases slightly for time-dependent activities as these activities introduce FP and FN instances in the activity inference model. For Activity-4, our proposed stage 4 activity inference HMM can achieve both accuracy and F-score over 95\%. The false positive rate (FPR) and false negative rate (FNR) are over 5\% and 3\% respectively for Activity-4. For Activity-4 and Activity-5, the accuracy of user activity inference decreases (93\% and 94\% respectively) while FPR and FNR increases. The reason for the increment of FPR and FNR is that different time-dependent user activities can have similar patterns over time with small changes in specific time instances. This affects the probability of occurring an activity calculated from HMM. In summary, an attacker can infer time-independent activities more accurately (with 100\% accuracy and F-score) than the time-dependent activities (with over 95\% accuracy and F-score).


Finally, note that 
an accurate user activity inference means that all the stages in the multi-stage attack have to be correctly guessed, which may lower the end-to-end successful inference rate of the attacker. For example, if the stage 1, 2, 3, and 4 are $X$, $Y$, $Z$, and $T$, respectively,  for an attacker, the probability of correctly guessing the Activity-4 of the user is $X\times Y \times Z \times T$. However, we also note that independently inferred information in every stage is also valuable as it may also include sensitive information (e.g., inferring the device type of a connected medical device may reveal the health status of the subject~\cite{gdpr, newaz2019healthguard}).  










\vspace{-6pt}
\section{Mitigating the Privacy Leaks} \label{sec:solution}

Despite the security vulnerabilities exploited before, as these privacy concerns are \textit{inherent} and \textit{insidious}, it is too hard to detect and avoid these types of threats associated with smart home devices. 
An attacker can passively listen to the wireless medium and record all the network traffic from a smart home environment without interrupting the normal activities of devices and their users.

\vspace{-5pt}

\subsection{Proposed Approach} 
In this sub-section, we propose a 
solution
based on generating
spoofed traffic.
In this way, 
even if the user is not at home, generating false activity for the user's presence traffic will mask the user's absence.

\begin{figure}
\centering
\begin{subfigure}{.5\columnwidth}
  \centering
  \includegraphics[width=\columnwidth]{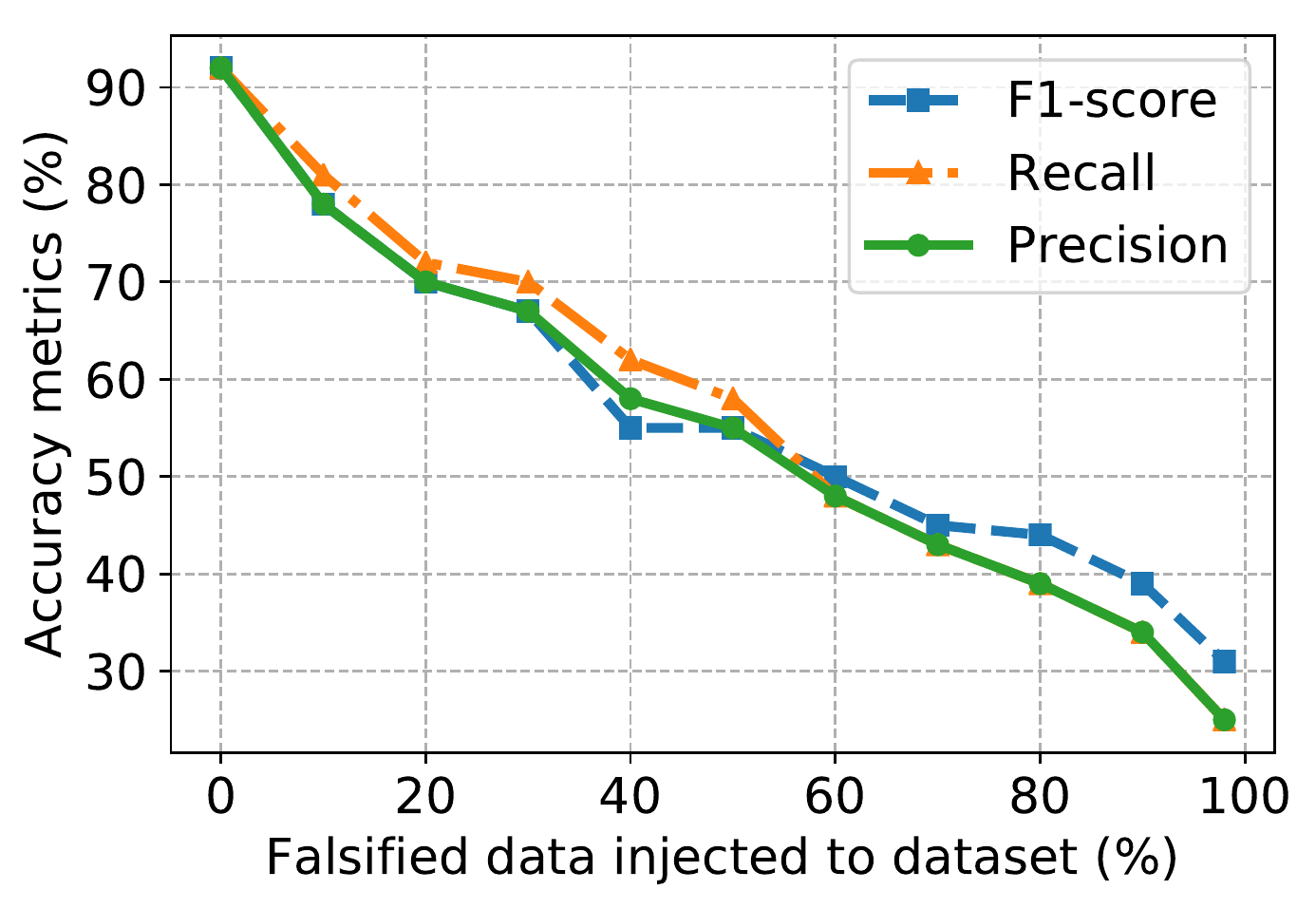}
   \caption{Device  state  detection}
   \label{fig:detection} 
\end{subfigure}%
\begin{subfigure}{.5\columnwidth}
  \centering
  \includegraphics[width=\columnwidth]{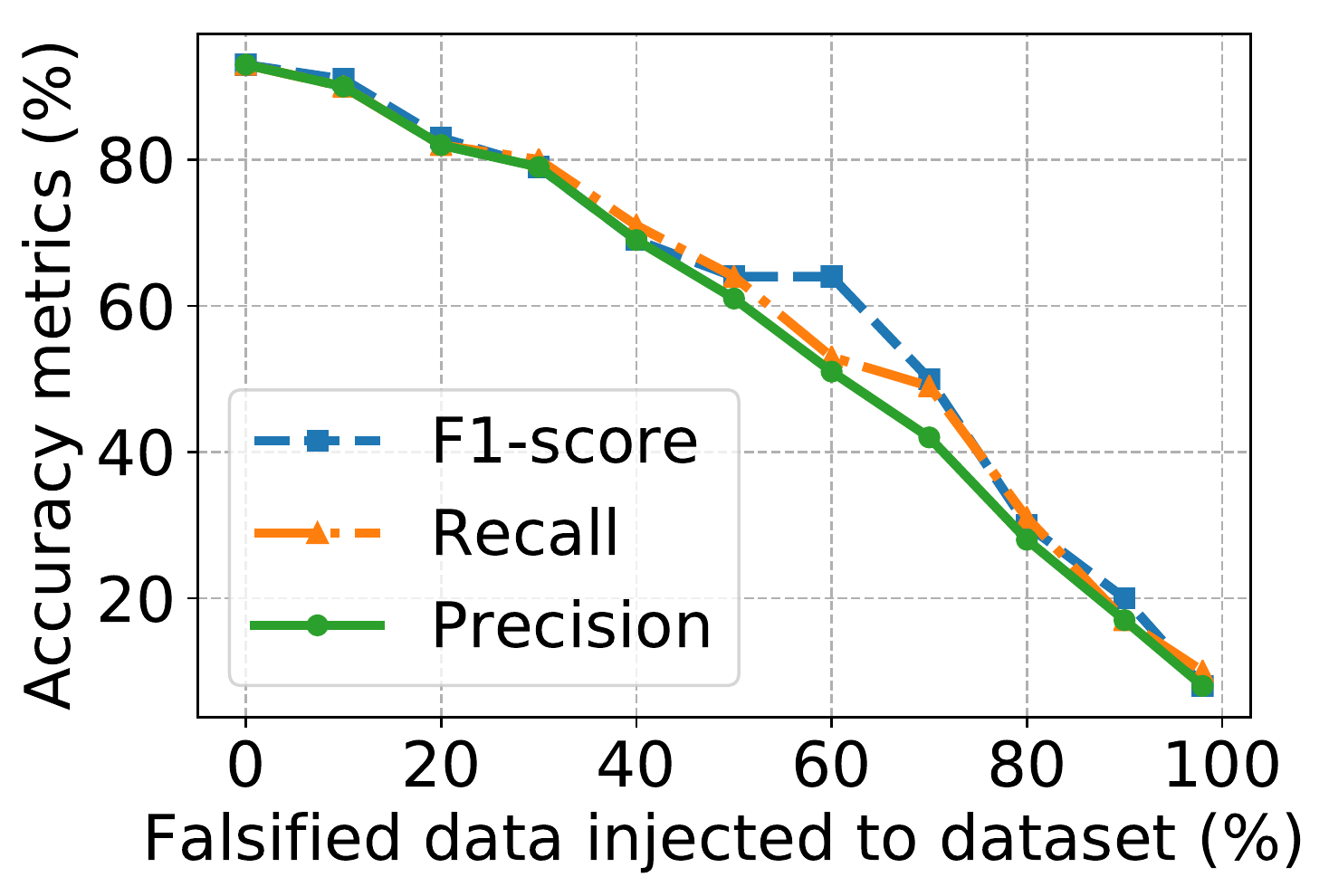}
   \caption{Device  state classification}
   \label{fig:classification}
\end{subfigure}
\caption{Impact  of  false  data  injection experiments on the attack accuracy.}
\label{fig:data_injection}
\end{figure}

In order to measure the efficacy of our proposed spoofed traffic, we investigated 
the injection of false packets by modifying the feature vectors and evaluated how the performance measurements would change. Then, we applied it to the device state detection and device activity classification attacks. Since the user activity inference is based on the results of the device state detection and device activity classification attacks, if we can falsify their results, the attacker will not able to infer the activities correctly. Particularly, we conducted a set of experiments where we injected falsified data into the training set to observe how the previously shown detection and classification algorithms would behave in such a situation. The results are shown in Figure~\ref{fig:data_injection}.

\noindent \textbf{Impact of False Data Injection on Device State Detection.}
Figure~\ref{fig:detection} shows the average of the accuracy measures for the kNN algorithm after increasingly injecting false packets. When there is no injected false packet, all of the devices have 91\% F1 score, then
it linearly 
decreases with the increase 
of false packets. For example, injecting false data equivalent to 10\% of packets exchanged 
during the observation time resulted in 
a decrease 
by 13\%. For 90\% false traffic addition, the accuracy of device state detection declined by about 57\%.
This shows that traffic injection can be efficiently used for hiding the state of devices from the adversary.

\vspace{5pt}

\noindent \textbf{Impact of False Data Injection on Device State Classification.} We injected the falsified data into the training data and computed the accuracy metrics in terms of F1 Score, Precision, and Recall. We injected 10\% falsified data and continued injecting until 90\% of the dataset contained false data. As can be seen in Figure~\ref{fig:classification}, the F1 Score plunges dramatically when injecting 90\% false data and reaches 15\%. This is due to the fact that randomly falsified features deteriorate traffic patterns used for classifying the devices' states. Also here, we can see that by injecting increasing amounts of fabricated traffic, the adversary can effectively be prevented from making inferences about the types of device events occurring. 

In real-life smart home environment, it is possible to introduce false/spoofed data packets (e.g., advertising, status/action request) by customizing the installed apps~\cite{berkay2018SaintTaintAnalysisUsenixSecurity} using the open source environments such as Samsung SmartThings. The installed apps can generate specific data packets at a specified interval without altering the real device states to trick the attacker from detecting device states. In our implementation, we only analyzed the effectiveness of injecting spoofed network traffic into the real traffic to hide the device states. An improved defense mechanism could use a more complex strategy to hide also the location at home as well as the source and destination of packets to avoid from device state detection and classification attacks. We leave this as an open problem to be explored in our future works.

\vspace{-7pt}
\section{Discussion} \label{sec:discussion}

\begin{figure}
  \centering
    \includegraphics[width=.75\columnwidth]{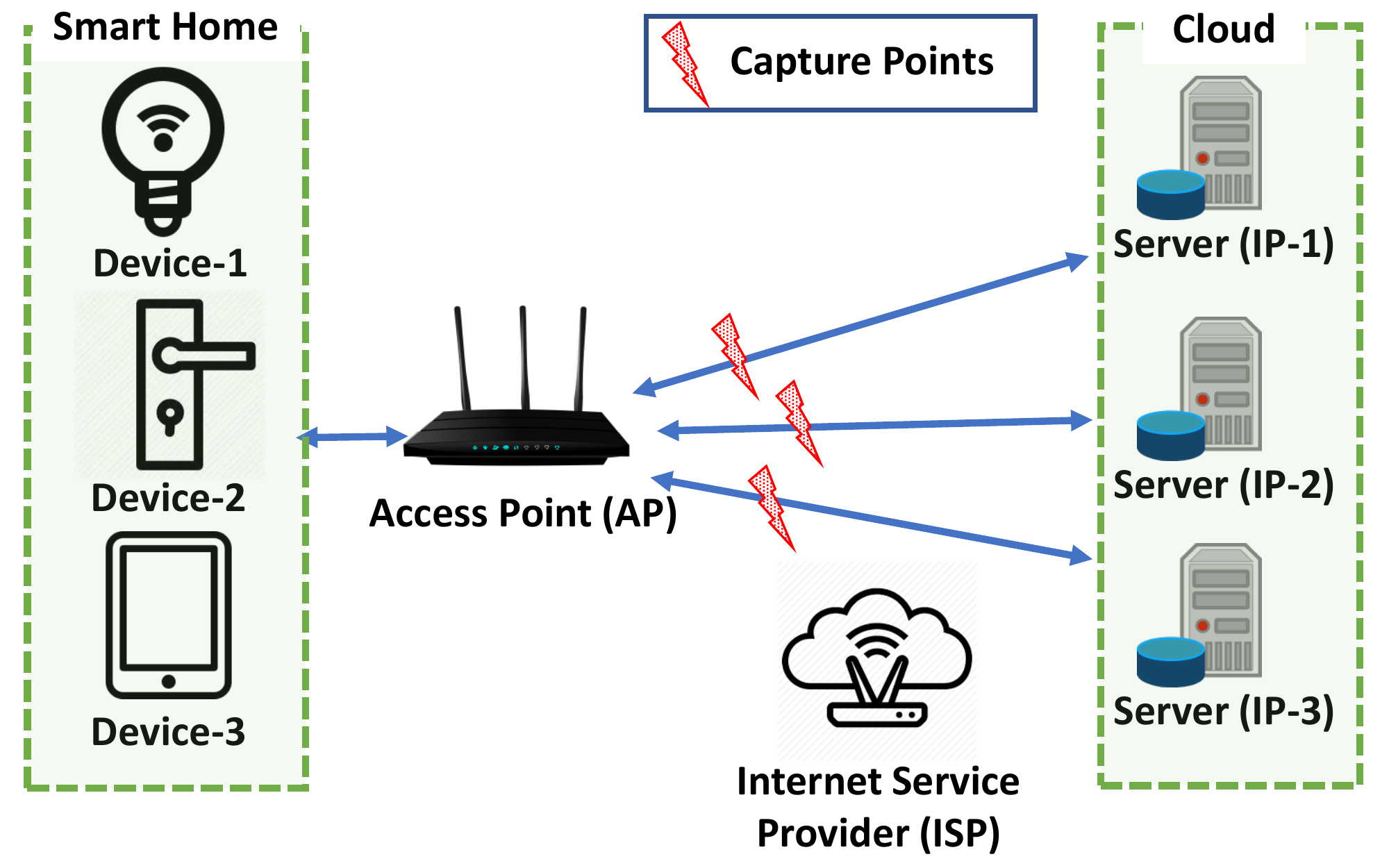}
    \vspace{-5pt}
     \caption{ Remote adversary model (e.g., a malicious ISP).}
     \vspace{-12pt}
     \label{fig:isp}
\end{figure}

\noindent \textbf{ISP as an adversary:}
Note that so far, our adversary model included  
only local adversary, where the adversary is within the range of radio frequency. An extension to this adversary model can be a remote adversary that can monitor outgoing network traffic of the smart home. A concrete example of such an adversary is an ISP.  Compared to the local adversary model considered 
in this paper, an ISP-like adversary has both advantages and disadvantages. It does not have to be within a range and it can see the source and destination IPs of the packets, which a local adversary can not see if the WPA encryption is enabled.
However, it can only collect the outgoing network traffic, not the internal two-way (upstream and downstream) 
network traffic as all the traffic is merged by the gateway (i.e., access point).  
Figure~\ref{fig:isp} shows the complete topology of the network from device to cloud.

As can be seen in Figure~\ref{fig:isp}, an ISP will only see the router's (i.e., gateway/access point)  MAC address. Therefore, it can not use the MAC addresses of the smart home devices for the device identification.  However, it can still try to use IPs in order to identify the devices and infer activities. Though there are number of challenges that attacker needs to solve in order to able use IP as a device identifier. First of all, if Network Address Translation (NAT) is deployed by the AP\footnote{Assuming IPv4 is still in use.}, the ISP can not find out the topology of the smart home and the number of devices. Even though NAT is 
not enabled, ZigBee and BLE devices have never been assigned an IP as they communicate with the AP through a hub, where only the hub they are connected to gets an IP.  Moreover,
devices do not communicate with only one server. Instead,  
sometimes multiple devices use one server (i.e., destination IP) as in the Samsung ST Hub, or sometimes one device can use multiple servers~\cite{copos2016anybody,apthorpe2017spying}. Therefore, even though the ISP-like attacker has some advantages (i.e., seeing IPs) over the local adversary, there are additional challenges that it needs to solve to get the same attack working. We leave this kind of adversary out of scope for now and will be studied in a future work.


\noindent \textbf{Multi-user vs. single user:} Smart home devices support multiple authorized users where more than one user can control and change the settings of smart devices. Additionally, multiple users can perform different activities within the smart environment at a time. This can create some false positive and false negative cases in user activity inference using our proposed method. Nonetheless, an attacker can still infer the device type and devices states from the network traffic. Additionally, the attacker can also infer the presence of multiple users and the specific point of ongoing activities in multi-user smart home environment using the network traffic. Compared to a multi-user scenario, a single user smart home environment is more vulnerable to our proposed threat as it is easier to infer a single on-going user activity in the smart home.

\noindent \textbf{Local vs. remote control:} To improve the user control and convenience over smart devices, smart homes offer remote access control in addition to traditional local access. Our proposed threat model can guess both local and remote access from location feature of the captured network traffic. This is a serious threat to user privacy as attackers can detect when a user is changing the state of a specific device remotely and perform malicious activities. For example, an attacker can infer when a user is accessing the smart lock remotely, which may result in physical access to the home environment. 

\noindent \textbf{Smart device diversity:} Smart devices have no common network protocols. Indeed, some of them such as WiFi, ZigBee, and BLE 
are more popular than others. This makes it harder to sniff all the devices that the smart home user is using. 
In addition to the diversity of 
network protocols, smart home devices come with different computational resources, hardware types, capabilities, exchanged data format etc. All of these differences in smart devices make it very challenging to build a generic solution as well as an attack. 
However, 
with our automated multi-stage privacy-attack, we showed the feasibility of the attack with the most popular network protocols, which covers the most of the commercial devices. 

\noindent \textbf{Generalizability of the attack:} As we noted in the assumptions, the attacker we considered in this paper does not need exactly the same devices to train its attack model, but it needs exact brand and device type to get the results presented in this paper as we use the $<brand,device-type>$ pair to uniquely identify devices. In other words, we assume at the end of the device identification stage of our attack, the attacker knows $<brand,device-type>$ pair. However, this assumption weakens the attack model. An attacker who can infer the device type and does not need the same device with the same brand would be more realistic. In order to remove this assumption, the same device type 
with different brands should be used to train the models and to attack (i.e., testing). 
It would be interesting to train and test the attack models on the same device type 
with different brands, or the same brand with different device types. Moreover, it would be also interesting to test the affects of model numbers, device configurations, or firmware updates etc.


\vspace{-8pt}
\section{Related Work} \label{sec:related}

\noindent \textbf{Identification using the encrypted network traffic .} The meta-data (e.g. MAC, traffic rate) of encrypted network traffic triggers possible threats including unintentional disclosure of the content or user. There is an extensive literature in the identification of the content from the encrypted network traffic. For example, web page identification~\cite{sun2002statistical}, web user identification~\cite{liberatore2006inferring}, protocol identification~\cite{wright2006inferring} are some of the research on the identification using the encrypted traffic. Not only identification attacks, but also the countermeasures have been studied in several studies~\cite{dyer2012peek,cai2012touching}. 

\noindent  \textbf{Fingerprinting Methods.} In all the aforementioned studies, either statistical techniques~\cite{velan2015survey} or machine learning methods~\cite{7265055,gtid_uluagac} were used to infer different sensitive information about the user and the context. Even ML has been used for the task of identification such as user, device, or website identification, in none of these studies, the attacks are timing-based as we have in our work.

\noindent \textbf{IoT Fingerprinting.} So far, in all the aforementioned studies the results showed that the used methods are efficient and the threat is  real, but the threat was limited to the web and online privacy of the user. Now with the emergence of IoT, it has been extended to every part of our daily lives and, with this, threats and countermeasures have also evolved~\cite{babun2018iotdots,acar2020usable,sikder2019multi}. The number of studies on the IoT fingerprinting through the network traffic has been increasing every day. Many studies have investigated the device type identification problem, where it has been sometimes proposed for both attacking~\cite{miettinen2017iot,shahid2018iot,bihl2017optimization,dong2019your,salman2019machine} and improving the security of smart home platforms~\cite{oconnor2019homesnitch,oconnor2019blinded,Birnbach:2019:PPE:3319535.3354254}. Moreover, some other works~\cite{copos2016anybody,apthorpe2017smart,apthorpe2017spying,apthorpe2017spying-2,junges2019passive,trimananda2019pingpong,ren2019information} worked on the device activity (event) inference problem, where the phrases device activity inference and user activity inference sometimes have been used interchangeably. In our work, we refer to the device activity (event) as the activity inferred from only one device. Even though sometimes the device activity and user activity would be the same thing (e.g., "coffee maker is ON" is the same as "the user is making coffee"), sometimes information from multiple devices is needed to infer one user activity correctly (e.g., see Figure~\ref{context}). We differentiate those two types of activities and provide a more generalized activity types in the fourth stage of our attack by modeling the user activities using HMM in Section~\ref{subsec:activity}.

\noindent \textbf{Difference from existing work.} Our work differs from the aforementioned studies in several ways: First, we are proposing a comprehensive method of end-to-end attack to infer the on-going user activities in a cascaded manner, where the previous studies have focused on only one stage of the attack. Note that putting all the different attack mechanisms and executing them successfully is a non-trivial task. Second, we are proposing the use of HMM for user activity modeling, where the device activities from multiple devices have been used to infer user activities. 
Last but not least, for the analysis of our attack, we performed experiments using the devices with WiFi, ZigBee, and BLE, where most of the previous studies have focused only on one of those wireless protocols.

\vspace{-8pt}
\section{Conclusion} \label{sec:conclusion}
In this paper, we explored how encrypted 
traffic 
from a smart home environment can be used to infer sensitive information about smart devices and sensors. Specifically, we introduced a novel multi-stage privacy attack, which an attacker can exploit to automatically detect and identify particular types of devices,  their actions, states, and related user activities
by passively monitoring the 
traffic of smart home devices and sensors.
%
Our evaluation on an extensive list of off-the-shelf smart home devices, sensors, and real users 
showed that an attacker 
can achieve very high accuracy (above \%90) in all the attack types.  As opposed to to earlier straightforward activity identification approaches, the novel multi-stage privacy attack can perform 
detection and identification automatically, is device-type and protocol-agnostic, and does not require extensive background knowledge or specifications of analyzed protocols. 
%
Finally, we propose a new yet effective mitigation mechanism to hide the real activities of the users. The effectiveness of the multi-stage privacy attack raises serious privacy concerns for any smart environment equipped with smart devices and sensors including personal homes, residences, hotel rooms, offices of corporations or government agencies.  

\vspace{-7pt}
\section*{Acknowledgment}\label{sec:acknowledgment}
This work is partially supported by the US National Science Foundation (Awards: NSF-CAREER-CNS-1453647, NSF-1663051, NSF-1718116),
Cyber Florida's Capacity Building Program, 
the EU LOCARD Project (agreement H2020-SU-SEC-2018-2019-2020- 8327351), 
the EU COLLABS Project (ICT-2019-2 GA 871518),
the Deutsche Forschungsgemeinschaft (DFG, German Research Foundation) - SFB 1119 - 236615297, and
Cisco Systems under the IoT Crowdsense project, gift \# 2019-197831 (3696). The views expressed are those of the authors only, not of the funding agencies.

\bibliographystyle{ACM-Reference-Format}
\bibliography{sample-base}

\appendix

\vspace{-5pt}
\section{Performance Metrics} \label{sec:metrics}
To evaluate our proposed novel attack, we used seven different performance metrics: True Positive Rate (TPR), False Negative Rate (FNR), True Negative  Rate (TNR),  False  Positive  Rate (FPR),  Precision, Accuracy, and F1-score. These can be calculated using following equations:
\begin{equation}\label{tp}
\begin{tiny}
{TPR\ (Recall)} = \frac{TP}{TP+FN},
\end{tiny}
\end{equation}
\begin{equation}
{FNR} = \frac{FN}{TP+FN}, 
\end{equation}
\begin{equation}
{TNR} = \frac{TN}{TN+FP}, 
\end{equation}
\begin{equation}
{FPR} = \frac{FP}{TN+FP}, 
\end{equation}
\begin{equation}\label{pre}
Precision = TP  /(TP+FP), 
\end{equation}
\begin{equation} \label{acc}
{Accuracy} = \frac{TP+TN}{TP+TN+FP+FN},
\end{equation}
\begin{equation}\label{F1}
{F1-score} = \frac{2*TP*TN}{TP+TN}, 
\end{equation}
where $TP= True\ Positive$, $FP= False\ Positive$, $TN= True\ Negative$ and $FN=False\ Negative$.
\vspace{-5pt}
\section{Case Studies\label{sec:case_studies}}
 In this section, we show the feasibility and possibility of privacy leaks from encrypted network traffic of smart home devices. We show that an attacker who can sniff the network traffic of the devices can easily infer some simple information without using any advanced techniques.  We consider one device for each protocol: 
Wemo Insight Switch (WiFi), 
Samsung ST Outlet (ZigBee), and August Smart Lock (BLE). We analyze the raw network traffic of each device and see if it is really possible to extract information from the network traffic, specifically from data rate.

\begin{figure}[t]
\centering
\begin{subfigure}[b]{0.425\textwidth}
   \includegraphics[width=\linewidth, trim= 0.9cm 0.5cm 0cm 1cm]{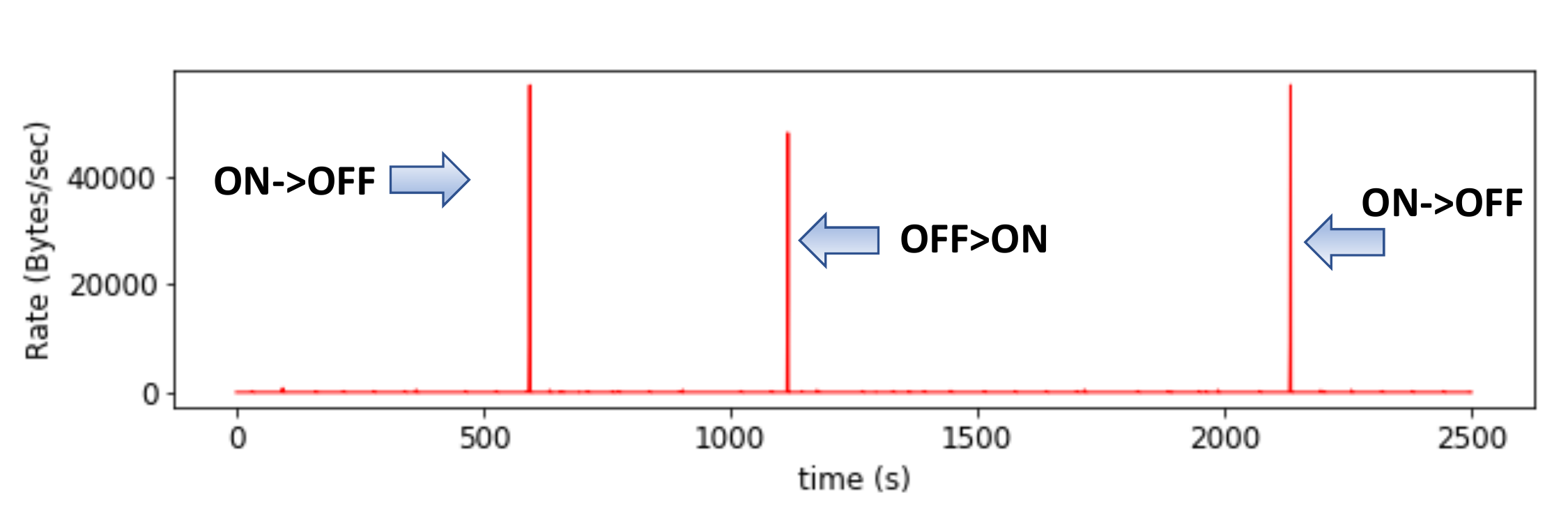}
   \caption{Wemo Insight Switch (WiFi)}
   \label{fig:switch} 
\end{subfigure}
\vspace{-3pt}
\begin{subfigure}[b]{0.425\textwidth}
   \includegraphics[width=1\linewidth,trim= 0.5cm 0.5cm 0cm 0cm]{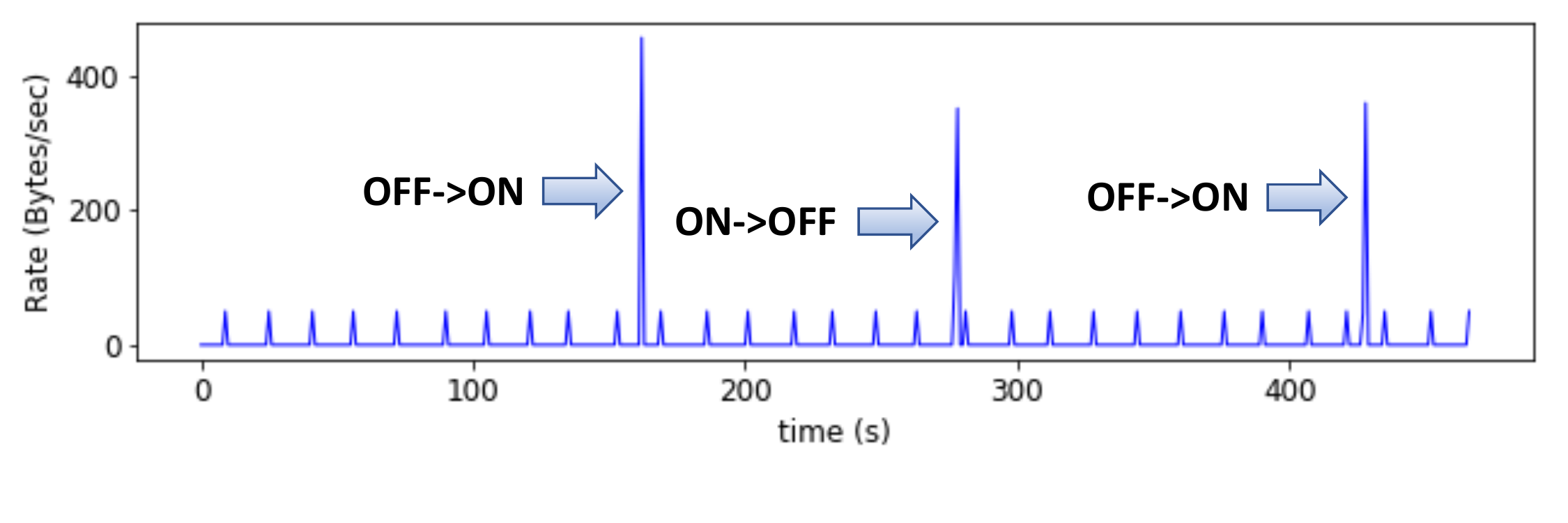}
   \caption{Samsung SmartThings Outlet (ZigBee)}
   \label{fig:outlet}
\end{subfigure}
\vspace{-3pt}
\begin{subfigure}[b]{0.425\textwidth}
   \includegraphics[width=1\linewidth,trim= 0.5cm 0.5cm 0cm 0cm]{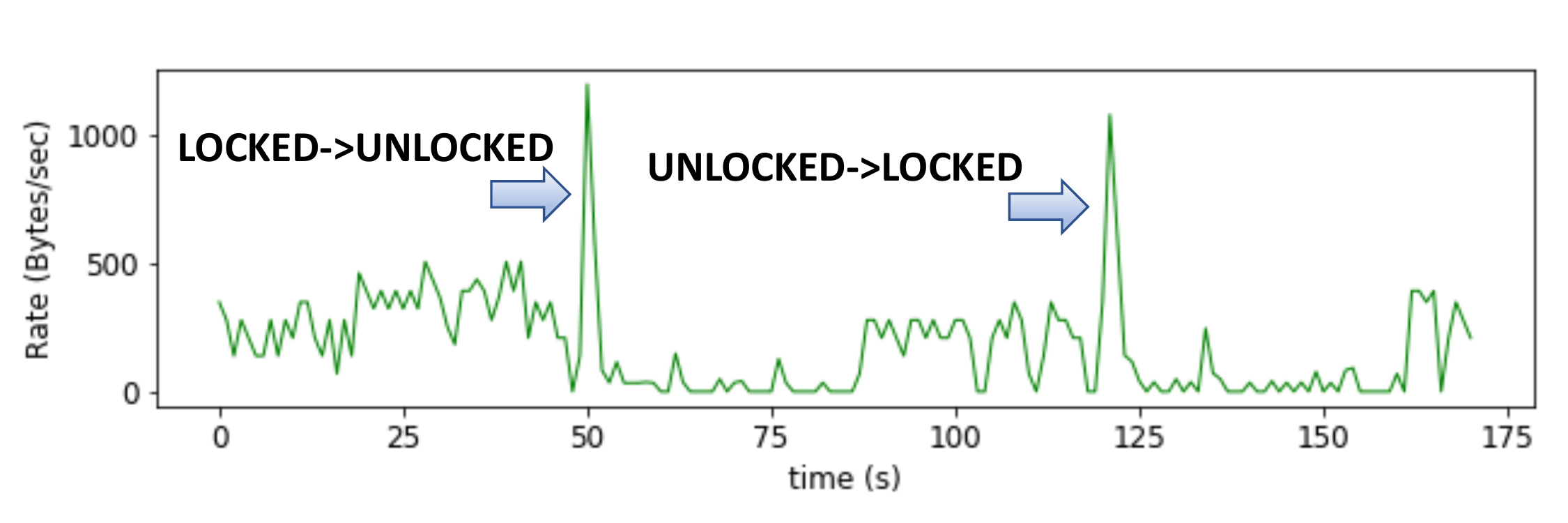}
   \caption{August Smart Lock (BLE)}
   \label{fig:lock}
\end{subfigure}
\caption{The traffic rates of (a) Wemo Insight Switch, (b) Samsung ST outlet, and (c) August Smart Lock. Here, a number of actions are illustrated, with many signals easily discerned by the naked eye. For instance, when the lock is turned on, the significant amount of packets are transmitted and received, which creates a peak in the traffic rate for a certain duration.\label{fig:all-3}}
\end{figure}


\subsubsection{Wemo Insight Switch (WiFi)}
Wemo Insight Switch is a Wifi-enabled device and used to monitor and control other appliances (e.g., smart light) from a smartphone. It has only two capabilities: ON and OFF. 
Figure~\ref{fig:switch} shows the data rate of the sample traffic collected from Wemo Insight Switch, where we illustrated a number of actions of the user to change the state of the device. As can be seen from the figure, the data rate shows a significant increase when the device state is changing. Therefore, the data rate clearly reveals the device state changes. In the first peak, the device's state is changed by the user, i.e., the device is turned on and in the second peak, the user turned off the device and so on.

\vspace{-5pt}

\subsubsection{Samsung ST Outlet (ZigBee)}
Samsung SmartThings (ST) Outlet uses ZigBee protocol to communicate with Samsung ST Hub. It can also act as a repeater and repeats the broadcast packet of Hub for the smart devices, which is not in the range of Hub. This increase the range of Hub. Other than repeating Hub's broadcasting packets, it has only two capabilities: ON and OFF. 
The traffic rate of a sample network capture of Samsung ST Outlet is plotted in Figure~\ref{fig:outlet}. In the given sample network traffic, the device's activity has been changed by the user three times, which clearly corresponds to the three large peaks. On the other hand, small peaks correspond to the repeating of the broadcast packets of the hub, which is periodic with 15 seconds. 
\vspace{-5pt}
\subsubsection{August Smart Lock (BLE)} 
The August Smart Lock communicates with the user's smartphone via BLE. In addition to locking and unlocking from the app on the smartphone, the owner (main user) can also give access to guest users through the web servers. The user can also enable the auto-unlock, where the lock is unlocked when the user is in range. However, the lock itself does not have the remote control capability. For remote access, it needs other accessories (e.g., WiFi bridge). Here, we only consider the BLE communication between the lock and smartphone.
Figure~\ref{fig:lock} shows the plot of the sample packet capture of August Smart Lock. As in the previous case studies, the transition between the device's actions can be clearly identified by the attacker. The small increase in the traffic rate in the first part of the capture is because of the advertising packets.

\end{document}